\documentclass[11pt]{article}
\usepackage[utf8]{inputenc}

\usepackage{epsfig,rotating,amsmath,lscape}
\usepackage{amssymb,latexsym,ifthen,bm,enumerate}
\usepackage{amsfonts}
\usepackage[numbers, sort&compress, super]{natbib}
\usepackage{color}
\usepackage{multirow}
\usepackage{alphalph}
\usepackage{wrapfig}
\usepackage{subcaption}
\usepackage{epstopdf}
\usepackage{csquotes}
\usepackage{enumitem}
\usepackage{comment}
\usepackage{todonotes}
\usepackage{adjustbox}
\usepackage{algorithm}
\usepackage{algpseudocode}
\usepackage{listings}
\usepackage[breaklinks]{hyperref}

\newcommand{\Var}{\operatorname{Var}}
\newcommand{\Cov}{\operatorname{Cov}}
\newcommand{\E}{\operatorname{E}}

\DeclareTextFontCommand{\emph}{\boldmath\bfseries}

\begin{document}
\title{Covariate adjustment and estimation of difference in proportions in randomized clinical trials}
\author{Jialuo Liu \and Dong Xi}
\date{%
    Gilead Sciences, Foster City, California, USA\\
    \date{}
}

\maketitle

\begin{abstract}
Difference in proportions is frequently used to measure treatment effect for binary outcomes in randomized clinical trials. The estimation of difference in proportions can be assisted by adjusting for prognostic baseline covariates to enhance precision and bolster statistical power. Standardization or G-computation is a widely used method for covariate adjustment in estimating unconditional difference in proportions, because of its robustness to model misspecification. Various inference methods have been proposed to quantify the uncertainty and confidence intervals based on large-sample theories. However, their performances under small sample sizes and model misspecification have not been comprehensively evaluated. We propose an alternative approach to estimate the unconditional variance of the standardization estimator based on the robust sandwich estimator to further enhance the finite sample performance. Extensive simulations are provided to demonstrate the performances of the proposed method, spanning a wide range of sample sizes, randomization ratios, and model misspecification. We apply the proposed method in a real data example to illustrate the practical utility.


\end{abstract}


\section{Introduction} \label{sec:intro}
In randomized clinical trials, statistical analyses with covariate adjustment can improve precision for estimating treatment effects \cite{food2023adjusting}. Here, covariates refer to prognostic variables that are measured at baseline, instead of those collected post-randomization. To adjust for these covariates, regression models for the outcome variable are often used, conditional on the randomized treatment assignments and specified covariates. Binary endpoints, such as clinical response, are often used as primary or secondary endpoints in clinical trials. Approaches have been proposed to analyze binary endpoints with covariate adjustment \cite{freedman2008randomization, tsiatis2008covariate, moore2009covariate, rosenblum2009using, ge2011covariate, daniel2021making, benkeser2021improving, morris2022planning, vanlancker2022combining, ye2023robust}.

Unlike continuous endpoints, for which analysis of covariance (ANCOVA) may be used to appropriately adjust for covariates, binary endpoints often require nonlinear regression models for analysis, such as the logistic regression. Covariate adjustment in these models may change the interpretation of the estimated treatment effect from an unconditional (or marginal) effect (i.e., an average effect over the distribution of covariates) to a conditional effect (i.e., an effect conditional on specific values of covariates). For frequently-used measures for binary endpoints such as the odds ratio, the unconditional effect may not be equal to the conditional effect or the average of conditional effects, which is often referred to as non-collapsibility \cite{gail1984biased, robins1986estimators}. There are also collapsible effect measures such as the difference in proportions and the ratio of proportions. In this article, we focus on unconditional estimators for binary outcomes, and in particular on the difference in proportions or the risk difference (used interchangeably in this article). Discussions have been provided on choices between conditional and unconditional estimators \cite{daniel2021making, morris2022planning}.

To obtain unconditional treatment effect estimates in the presence of covariate adjustment, a useful methodology known as standardization or g-computation can be used \cite{freedman2008randomization, ge2011covariate, food2023adjusting}. Here are the steps to obtain a standardization estimator for the difference in proportions of a binary outcome:
\begin{algorithm}
	\begin{algorithmic}[1]
		\State Fit a logistic regression with treatment assignments and covariates to the randomized data.
		\State For each subject, predict the model-based probability of response given the subject's covariates and under each treatment assignment, e.g., treatment and control separately.
		\State Average over the predicted probabilities for each treatment assignment.
		\State Contrast the average probabilities to obtain an unconditional estimator for the difference in proportions.
	\end{algorithmic}
\end{algorithm}

Standardization based on the logistic model has been shown to be robust to model misspecification \cite{freedman2008randomization, steingrimsson2017improving}. There are also other ways to obtain unconditional effects for binary outcomes such as the target maximum likelihood estimators \cite{moore2009covariate} and the semi-parametric estimators \cite{tsiatis2008covariate}. They are similar to standardization estimators for the difference of proportions using the logistics regression. In this article, we focus on the standardization approach for its simplicity.

While the standardization estimator can be obtained straightforwardly, common approaches for variance and confidence interval calculation include the delta method, efficient influence functions \cite{moore2009covariate, jann2023estimation}, and semi-parametric approaches \citep{ye2023robust}. For example, the delta method based on the model-based variance-covariance matrix has been proposed \cite{ge2011covariate}. Further, it is pointed out that this variance estimator may suffer from model misspecification and may underestimate the variance due to missing a non-zero component \cite{ye2023robust}. A robust variance estimator based on semi-parametric approaches is proposed \cite{ye2023robust}.

Based on these results in the literature, we plan to derive the variance estimator for standardization of the risk difference by using the delta method. As suggested by the FDA guideline on covariate adjustment \cite{food2023adjusting}, the Huber-White robust ``sandwich" estimator could be more robust to model misspecification \cite{rosenblum2009using}. However, there are many versions of the sandwich estimator, for example in the R {\tt sandwich} package \cite{zeileis2020various}, and it is not clear which one is more appropriate than others. Further, the variance estimator conditioning on covariates which may underestimate the unconditional variance \cite{ge2011covariate}. We try to complete the formulation with the additional adjustment to avoid underestimation and obtain an unconditional variance estimator.

In Section \ref{sec:estimation}, we propose an unconditional variance estimator for standardization of the risk difference based on the delta method. In Section \ref{sec:simulation}, we conduct an extensive simulation study to compare the performance of our proposed variance estimator against various alternatives across a wide range of sample sizes, under both correct and wrong model specifications. In Section \ref{sec:case}, we apply proposed approaches in a case study using existing R packages to illustrate the simple implementation. In Section \ref{sec:conclusion}, we conclude with conclusions and discussions.

\section{Standardization estimator and its variances} \label{sec:estimation}

\subsection{Standardization estimator} \label{sec:standardization}
Consider a potential outcome $Y^{(z)}$ for a patient were they given treatment $z$, possibly contrary to the actual treatment assignment. Let $Y^{(z)}$ be a binary outcome, indicating whether a patient experienced a certain clinical outcome $\left(Y^{(z)}=1\right)$ or not $\left(Y^{(z)}=0\right)$. To simplify the discussion, we consider a binary randomized treatment assignment $Z$, where $Z=1$ represents the assignment of treatment and $Z=0$ represents control. Clinical trials often collect baseline covariates $\bm{W}$ which are related to the outcome. Utilizing such covariates, we focus on the inference for the average treatment effect which is the difference in proportions or the risk difference (RD) for binary variables as
\begin{align}\label{eq:rd}
	\mathrm{RD} &= \E\left(Y^{(1)}\right) - \E\left(Y^{(0)}\right) = \Pr\left(Y^{(1)}=1\right)-\Pr\left(Y^{(0)}=1\right).
\end{align}

Assume there are $n$ subjects enrolled in a randomized trial. From the $i$-th subject, we observe the randomized treatment assignment $z_i$, baseline covariates $\bm{w}_i$, and the binary outcome of interest $y_i$ for $i=1, \ldots, n$. A popular statistical model to analyze binary outcomes is the logistic regression, which can be incorporated in standardization to obtain the unconditional estimate of RD as in Algorithm \ref{alg:standardization}. 
\begin{algorithm}
	\caption{Standardization (G-computation) with the logistic regression}\label{alg:standardization}
	\begin{algorithmic}[1]
		\State Fit a logistic regression model such that 
		\begin{equation}\label{eq:logistic}
			\mathrm{logit}\left\{\Pr(Y=1|Z,\bm{W})\right\}=\log\left\{\frac{\Pr(Y=1|Z,\bm{W})}{1-\Pr(Y=1|Z,\bm{W})}\right\} = \beta_0 + \beta_1 Z + \bm{\beta}_2^\top \bm{W},
		\end{equation}
		where $\beta_0$, $\beta_1$, and $\bm{\beta}_2$ are coefficients to be estimated from the data, and their maximum likelihood estimators are $b_0$, $b_1$, and $\bm{b}_2$, respectively.
		\State Use the fitted logistic regression model in \eqref{eq:logistic} to predict the probability of response $\Pr(y_i=1|z_i=1,\bm{w}_i)$ and $\Pr(y_i=1|z_i=0,\bm{w}_i)$ for each subject $i$ as if they had been assigned to the treatment or control group, respectively. We obtain $\widehat{\Pr}\left(y_i=1|z_i=1,\bm{w}_i\right)=\mathrm{logit}^{-1}\left(b_0+b_1+\bm{b}_2^\top\bm{w}_i\right)$, and $\widehat{\Pr}\left(y_i=1|z_i=0,\bm{w}_i\right)=\mathrm{logit}^{-1}\left(b_0+\bm{b}_2^\top\bm{w}_i\right)$, where $\mathrm{logit}^{-1}(\cdot)=\frac{\exp{(\cdot)}}{1+\exp{(\cdot)}}$ is the inverse of the logit function defined in \eqref{eq:logistic}.
		\State Average over the entire sample to obtain the average response rate for the treatment and control group by $\frac{1}{n}\sum_{i=1}^n \widehat{\Pr}\left(y_i=1|z_i=1,\bm{w}_i\right)$ and $\frac{1}{n}\sum_{i=1}^n \widehat{\Pr}\left(y_i=1|z_i=0,\bm{w}_i\right)$, respectively.
		\State The unconditional treatment effect estimate of RD is given by
		\begin{equation}\label{eq:rd_hat}
			\widehat{\mathrm{RD}} = \frac{1}{n}\sum_{i=1}^n \widehat{\Pr}(y_i=1|z_i=1,\bm{w}_i)  - \frac{1}{n}\sum_{i=1}^n \widehat{\Pr}(y_i=1|z_i=0,\bm{w}_i).
		\end{equation}
	\end{algorithmic}
\end{algorithm}

The logistic regression model in \eqref{eq:logistic} can be replaced by other choices. For example, we could estimate RD without covariate adjustment by fitting a logistic regression model with only the treatment assignment as 
\begin{equation}\label{eq:logistic_ncovariate}
	\mathrm{logit}\left\{\Pr(Y=1|Z)\right\} = \beta_0 + \beta_1 Z.
\end{equation}
In addition, the logistic regression model in \eqref{eq:logistic} can be replaced by a more sophisticated model, e.g., a logistic regression with treatment by covariate interactions \citep{moore2009covariate}, or two separate regression models for treatment and control respectively \citep{tsiatis2008covariate}. Using these more sophisticated models may improve efficiency, but they usually require a larger sample size due to more parameters to estimate, compared to a simpler model like \eqref{eq:logistic}. In this article, we focus the exploration for smaller sample sizes, which usually are feasible only for simpler models. For ease of illustration, interpretation and computation, we choose the logistic regression model in \eqref{eq:logistic}. Standardization based on this model has been shown to be robust to model misspecification \citep{steingrimsson2017improving}. Therefore, the focus of this article is on the variance estimation for standardization of RD based on the logistic regression in \eqref{eq:logistic}.

\subsection{Variance estimation} \label{sec:delta}
We now consider variance estimators of the standardization estimator in Algorithm \ref{alg:standardization}. A variance estimator using the delta method has been proposed \cite{ge2011covariate}, which conditions on the observed covariate $\bm{W}$ and $Z$. Let $X = (\bm{x}_{1},\ldots,\bm{x}_{n})^\top$ be the design matrix with $\bm{x}_{i}=(1,z_i,\bm{w}_i^\top)^\top, i=1,\ldots,n$. Let $\bm{b} = (b_0,b_1,\bm{b}_2^\top)^\top$ denote the vector of the maximum likelihood estimators in the logistic regression \eqref{eq:logistic}. The predicted probability for the $i$-th subject under the assigned treatment $z_i$ is $\widehat{\pi}_i=\mathrm{logit}^{-1}\left(\bm{x}_{i}^{\top} \bm{b}\right)$. To predict the probability for the $i$-th subject under counterfactual treatment assignments, denote column vectors under treatment and control respectively as $\bm{x}_{i(1)}=(1,1,\bm{w}_i^\top)^\top$ and $\bm{x}_{i(0)}=(1,0,\bm{w}_i^\top)^\top$. The predicted probability under the counterfactual treatment assignments is $\widehat{\pi}_{i(1)}=\mathrm{logit}^{-1}\left(\bm{x}_{i(1)}^{\top} \bm{b}\right)$ and $\widehat{\pi}_{i(0)}=\mathrm{logit}^{-1}\left(\bm{x}_{i(0)}^{\top}\bm{b}\right)$, respectively under treatment and control. Define $\widehat{\pi}_{(1)} = \frac{1}{n}\sum_{i=1}^{n} \widehat{\pi}_{i(1)}$ and $\widehat{\pi}_{(0)} = \frac{1}{n}\sum_{i=1}^{n} \widehat{\pi}_{i(0)}$ as the average probabilities under counterfactual treatment assignments.

By using the delta method, the conditional variance of the standardization estimator for RD can be estimated as follows. For $j=0,1$, we have
$$
\frac{\partial\widehat{\pi}_{(j)}}{\partial \bm{b}} = \frac{1}{n}\sum_{i=1}^{n} \frac{\partial \mathrm{logit}^{-1}\left(\bm{x}_{i(j)}^{\top}\bm{b}\right)}{\partial \bm{b}}= \frac{1}{n}\sum_{i=1}^{n} \bm{x}_{i(j)}\widehat{\pi}_{i(j)} \left(1-\widehat{\pi}_{i(j)}\right).
$$
The conditional variance estimator of RD has the following form:
\begin{equation}\label{eq:delta}
	\left(\frac{\partial\widehat{\pi}_{(1)}}{\partial \bm{b}} - \frac{\partial\widehat{\pi}_{(0)}}{\partial \bm{b}} \right)^{\top} V \left(\frac{\partial\widehat{\pi}_{(1)}}{\partial \bm{b}} - \frac{\partial\widehat{\pi}_{(0)}}{\partial \bm{b}}\right),
\end{equation}
where $V$ is the estimated variance-covariance matrix of $\bm{b}$.

There are multiple choices of $V$ that can be plugged in. The model-based estimator of $V_{\text{model}}$ from the logistic regression model is one option \cite{ge2011covariate}. While this is easy to obtain, it may be subject to biases under model misspecification \cite{ye2023robust}. An alternative choice, is to use a robust sandwich estimator defined as $V_{\text{sandwich}}$ \cite{food2023adjusting}. The core idea for the sandwich estimator is to further adjust from $V_{\text{model}}$ for residuals. A popular choice of the sandwich estimator is
\begin{equation}\label{eq:hc}
	V_{\text{sandwich}}=V_{\text{model}}\left[X^{\top}\text{diag}\left\{\frac{\widehat{\epsilon_i}^2}{(1-h_i)^\delta}\right\} X\right]V_{\text{model}},
\end{equation}
where $\widehat{\epsilon}_i = y_i - \widehat{\pi}_i$ is the $i$-th residual, and $h_i$ is the $i$-th diagonal element of the hat matrix $W^{\frac{1}{2}}X(X^{\top}WX)^{-1}X^{\top}W^{\frac{1}{2}}$, where $W$ is a diagonal matrix whose $i$-th element is $\widehat{\pi}_i(1-\widehat{\pi}_i)$. There are two choices of $\delta$ corresponding to two popular choices of the sandwich estimator. When $\delta=1$, \eqref{eq:hc} is called the version HC2 \cite{mackinnon1985some, horn1975estimating}; when $\delta=2$, \eqref{eq:hc} is called the version HC3 \cite{long1998correcting}.

The conditional variance estimator in \eqref{eq:delta} may neglect the variability in the covariate space and thus underestimate the unconditional variance \cite{ye2023robust}. Therefore, we propose an unconditional variance estimator as follows. By the law of total variance, the unconditional variance can be decomposed to two parts:
\begin{equation} \label{eq:ltv}
	\Var\left(\widehat{\mathrm{RD}}\right) =\E\left\{\Var\left(\widehat{\mathrm{RD}}|X\right)\right\}+\Var\left\{\E\left(\widehat{\mathrm{RD}}|X\right)\right\},
\end{equation}
The first term in \eqref{eq:ltv} is the conditional variance which can be estimated by \eqref{eq:delta}. The second term in \eqref{eq:ltv} carries additional variability from the covariates. It can be estimated as the sample variance of the mean of $\widehat{\pi}_{i(1)}-\widehat{\pi}_{i(0)}$, i.e., the square of the standard error of RD. Let the sample variance of $\widehat{\pi}_{i(1)}-\widehat{\pi}_{i(0)}$ be $\widehat{\sigma}^2_{\mathrm{RD}}$. Then $\Var\left\{\E\left(\widehat{\mathrm{RD}}|X\right)\right\}$ can be estimated as $\widehat{\sigma}^2_{\mathrm{RD}}/n$. Therefore, the proposed variance estimator is
\begin{equation}\label{eq:impvar}
	\Var\left(\widehat{\mathrm{RD}}\right)= \left(\frac{\partial\widehat{\pi}_{(1)}}{\partial \bm{b}} - \frac{\partial\widehat{\pi}_{(0)}}{\partial \bm{b}} \right)^{\top} V_{\text{sandwich}} \left(\frac{\partial\widehat{\pi}_{(1)}}{\partial \bm{b}} - \frac{\partial\widehat{\pi}_{(0)}}{\partial \bm{b}}\right) + \widehat{\sigma}^2_{\mathrm{RD}}/n.
\end{equation}
One advantage of this proposed variance estimator lies in implementation by using existing R packages {\tt margins} \cite{leeper2021margins} and {\tt sandwich} \cite{zeileis2020various}, as illustrated later in Section \ref{sec:case}. In the next section, we provide a simulation study to compared its performance against other methods in the literature.

\section{Simulation}\label{sec:simulation}
We conduct a simulation study to compare the performance of five different families of variance estimation methods with a total of nine methods. The first four families (including seven methods) adjust for covariates and the fifth family does not adjust for covariates. R programs to implement all methods are provided at \url{https://github.com/jialuo-liu/covadj}.
\begin{itemize}
	\item The first family uses the delta method to estimate the conditional variance which is the first term in \eqref{eq:ltv}. They include (M1) the delta method using the model-based variance estimator (Delta (Model)) \cite{ge2011covariate}, (M2) the delta method using the sandwich variance estimator HC2 (Delta (HC2)), (M3) the delta method using the sandwich variance estimator HC3 (Delta (HC3)). These methods can be implemented using the {\tt margins} R package \cite{leeper2021margins} and the {\tt sandwich} R package \cite{zeileis2020various}, as suggested previously \cite{betz2023tutorials}.
	\item The second family includes one method (M4) using the efficient influence function (EIF) \citep{van2011targeted}. Details are provided in Appendix \ref{sec:eif}.
	\item The third family includes one method (M5) using the semi-parametric variance estimator (Semi-parametric), which can be implemented using the {\tt RobinCar} R package \citep{ye2023robust}. Details are provided in Appendix \ref{sec:ye}.
	\item The fourth family includes our proposed unconditional variance estimator with (M6) the HC2 sandwich estimator in \eqref{eq:impvar} and (M7) the HC3 sandwich estimator in \eqref{eq:impvar}. We focus on these two versions of the sandwich estimator in the main body of this article, and provide a more detailed investigation of all versions in the {\tt sandwich} R package \cite{zeileis2020various} in Appendix \ref{sec:allhc}. These methods can be implemented using existing R packages such as the {\tt margins} package \cite{leeper2021margins}. Code examples are provided in Section \ref{sec:case}.
	\item The fifth and the last family includes methods which do not adjust for covariates using standardization in \eqref{eq:logistic_ncovariate} and the proposed variance estimator in \eqref{eq:impvar}. This family includes (M8) using the HC2 sandwich estimator (Unadjusted (HC2)) and (M9) using the HC3 sandwich estimator (Unadjusted (HC3)). 
\end{itemize}

We consider three independent variables including the randomized treatment assignment $Z$, a baseline continuous covariate $X_\mathrm{cont}$ and a baseline binary covariate $X_\mathrm{cat}$. The continuous covariate $X_\mathrm{cont}$ is generated from a standard normal distribution, while the binary covariate $X_\mathrm{cat}$ follows a Bernoulli distribution with a probability of 0.5. We also consider two randomization ratios, 1:1 and 2:1, using the stratified simple randomization by the binary covariate $X_\mathrm{cat}$. We consider a wide range of total sample sizes 30, 60, 90, 150, 360, and 900. Factoring in the two randomization ratios of 2:1 and 1:1, we investigate the following minimum per-group sample sizes: 10 and 15, 20 and 30, 30 and 45, 50 and 75, 120 and 180, 300 and 450, respectively. In total, we evaluate $12$ sample size scenarios, with 100,000 simulations for each scenario. We also investigate other randomization schemes including the simple randomization, whose results are provided in the Supplementary Materials.
For a better presentation of results, we separate the sample sizes of 150, 360, 900 from the sample sizes of 30, 60, 90, where the former is focused on relatively large sample situations and the latter is more on relatively small samples.

We evaluate the performance of the variance estimation methods using several operating characteristics, including the average of the standard errors, the coverage probability of the 95\% confidence interval, and the probability of rejecting the null hypothesis of no treatment effect at a two-sided significance level of 0.05. The 95\% confidence interval for $\widehat{\mathrm{RD}}$ is 
\begin{equation}\label{eq:ci}
	\widehat{\mathrm{RD}} \pm z_{1-\alpha/2}\sqrt{\Var\left(\widehat{\mathrm{RD}}\right)},
\end{equation}
where $z_{1-\alpha/2}$ is the $\left(1-\alpha/2\right)\times 100$-th percentile of the standard normal distribution. Note that all methods with covariate adjustment (M1-M7) share the same point estimate of RD. Thus any disparities in their performance are solely attributed to the difference in variance estimation methods.

In all simulations, the logistic regression model fitted in standardization includes the intercept, the main term of treatment assignment $Z$, the continuous covariate $X_\mathrm{cont}$, and the binary covariate $X_\mathrm{cat}$. Non-convergent model fitting may occur especially with smaller sample sizes or rare events. In such cases, we adopt an iterative approach by removing one covariate at a time until convergence is reached. If the removal of all covariates becomes necessary, an unadjusted approach \eqref{eq:logistic_ncovariate} is used. To ensure a fair comparison, this strategy is applied to all methods involving covariate adjustment to handle non-convergence.

\subsection{Correct model specification}\label{sec:simcorrect}
The binary response variable $Y$ is generated from a logistic regression model:
$$
\mathrm{logit}\left\{P(Y=1|Z, X_\mathrm{cont}, X_\mathrm{cat})\right\}=\beta_0+\beta_1 Z+\beta_2  X_\mathrm{cont} + \beta_3 X_\mathrm{cat},
$$
which has the same form as the logistic regression model fitted in standardization in \eqref{eq:logistic}. We consider three scenarios of true parameters in Table \ref{tab:scenarios}. Scenarios 1 and 2 represent the cases with positive treatment effect ($\beta_1>0$), where Scenario 2 has a larger coefficient for the treatment assignment than Scenario 1. Scenario 3 represents the null hypothesis, where there is no treatment effect ($\beta_1=0$).

\begin{table}[!ht]
	\centering
	\begin{tabular}{ccccc}
		\hline
		Scenario & $(\beta_0,\beta_1,\beta_2,\beta_3)$ & $\pi_{(0)}$ & $\pi_{(1)}$ & RD \\ 
		\hline
		1 & $(-1.7, 1.1,  3.0, -3.0)$ & $0.20$ &$0.29$  & $0.09$   \\
		2 & $(-4.0, 2.0,  4.2, -3.0)$ & $0.13$ &$0.23$  & $0.11$   \\
		3 & $(-1.2, 0.0,  1.0, -1.0)$ & $0.20$ &$0.20$  & $0.00$   \\
		\hline
	\end{tabular}
	\caption{True parameters for simulation under correct model specification.} \label{tab:scenarios}
\end{table}

We first consider the standard error and the coverage probability of the 95\% confidence interval for larger samples with the total sample sizes of 150, 360, and 900. Table~\ref{tab:secoverH} provides these results for Scenarios 1 and 2 under the randomization ratios 1:1 and 2:1. As the total sample size increases, the standard error decreases for all methods under all scenarios. Methods M1-M7 have similar standard errors which are smaller than those from M8-M9. This reflects the efficiency gain by including prognostic covariates $X_\mathrm{cat}$ and $X_\mathrm{cont}$ in M1-M7 compared to no covariate adjustment in M8-M9.

\begin{table}[!ht]
	\centering
	\noindent\adjustbox{max width=\textwidth}{%
		\begin{tabular}{lccc|ccc}
			\hline
			& \multicolumn{3}{c}{Standard error} & \multicolumn{3}{c}{Coverage probability}\\ \cline{2-4} \cline{5-7}
			\multicolumn{1}{r}{Total sample size:} & 150 & 360 & 900 & 150 & 360 & 900 \\
			\multicolumn{1}{r}{Randomization ratio:} & 1:1 (2:1) & 1:1 (2:1) & 1:1 (2:1) & 1:1 (2:1) & 1:1 (2:1) & 1:1 (2:1) \\\hline
			\multicolumn{1}{l|}{Method} & \multicolumn{6}{c}{Scenario 1: Moderate treatment effect} \\\hline
			M1: Delta (model)         & 0.046 (0.048) & 0.030 (0.031) & 0.019 (0.020) & 0.934 (0.935) & 0.943 (0.943) & 0.947 (0.947) \\ 
			M2: Delta (HC2)           & 0.047 (0.049) & 0.030 (0.032) & 0.019 (0.020) & 0.941 (0.939) & 0.946 (0.945) & 0.948 (0.947) \\ 
			M3: Delta (HC3)           & 0.049 (0.051) & 0.031 (0.032) & 0.019 (0.020) & 0.948 (0.947) & 0.949 (0.948) & 0.949 (0.948) \\ 
			M4: EIF                   & 0.047 (0.049) & 0.030 (0.032) & 0.019 (0.020) & 0.942 (0.939) & 0.948 (0.947) & 0.950 (0.950) \\ 
			M5: Semi-parametric       & 0.048 (0.050) & 0.031 (0.032) & 0.019 (0.020) & 0.947 (0.943) & 0.950 (0.949) & 0.951 (0.951) \\ 
			M6: Proposed (HC2)        & 0.048 (0.050) & 0.031 (0.032) & 0.019 (0.020) & 0.948 (0.946) & 0.950 (0.950) & 0.951 (0.951) \\ 
			M7: Proposed (HC3)        & 0.050 (0.052) & 0.031 (0.032) & 0.019 (0.020) & 0.954 (0.953) & 0.953 (0.953) & 0.952 (0.952) \\ 
			M8: Unadjusted (HC2)      & 0.070 (0.073) & 0.045 (0.047) & 0.029 (0.030) & 0.949 (0.946) & 0.950 (0.948) & 0.950 (0.949) \\ 
			M9: Unadjusted (HC3)      & 0.071 (0.073) & 0.045 (0.047) & 0.029 (0.030) & 0.950 (0.948) & 0.950 (0.949) & 0.950 (0.950) \\ \hline
			\multicolumn{1}{l|}{Method} & \multicolumn{6}{c}{Scenario 2: Large treatment effect} \\\hline
			M1: Delta (model)         & 0.037 (0.038) & 0.024 (0.025) & 0.015 (0.016) & 0.922 (0.922) & 0.931 (0.932) & 0.935 (0.936) \\ 
			M2: Delta (HC2)           & 0.039 (0.040) & 0.025 (0.025) & 0.015 (0.016) & 0.932 (0.926) & 0.935 (0.934) & 0.936 (0.937) \\ 
			M3: Delta (HC3)           & 0.041 (0.042) & 0.025 (0.026) & 0.016 (0.016) & 0.944 (0.940) & 0.940 (0.939) & 0.938 (0.939) \\ 
			M4: EIF                   & 0.040 (0.040) & 0.026 (0.026) & 0.016 (0.017) & 0.940 (0.934) & 0.946 (0.945) & 0.948 (0.948) \\ 
			M5: Semi-parametric       & 0.040 (0.041) & 0.026 (0.026) & 0.016 (0.017) & 0.943 (0.938) & 0.947 (0.946) & 0.949 (0.948) \\ 
			M6: Proposed (HC2)        & 0.041 (0.042) & 0.026 (0.027) & 0.016 (0.017) & 0.949 (0.945) & 0.949 (0.949) & 0.949 (0.949) \\ 
			M7: Proposed (HC3)        & 0.043 (0.044) & 0.026 (0.027) & 0.016 (0.017) & 0.957 (0.955) & 0.953 (0.953) & 0.951 (0.951) \\ 
			M8: Unadjusted (HC2)      & 0.062 (0.063) & 0.040 (0.041) & 0.025 (0.026) & 0.949 (0.945) & 0.951 (0.949) & 0.949 (0.948) \\ 
			M9: Unadjusted (HC3)      & 0.063 (0.064) & 0.040 (0.041) & 0.025 (0.026) & 0.950 (0.947) & 0.951 (0.950) & 0.950 (0.949) \\   \hline
	\end{tabular}}
	\caption{Standard error and coverage probability of the 95\% confidence interval under 1:1 (2:1) stratified randomization and correct model specification for the total sample size of 150, 360, and 900.}
	\label{tab:secoverH}
\end{table}

In terms of the coverage probability of the 95\% confidence interval for the total sample sizes of 150, 360, and 900, methods estimating the conditional variance (M1-M3) show slight undercoverage for Scenario 1 and a more pronounced undercoverage for Scenario 2. Substituting the sandwich estimators of HC2 (M2) and HC3 (M3) for the model-based variance (M1) improves the coverage. The methods using EIF (M4) and the semi-parametric approach (M5) have a small undercoverage for Scenario 2. The proposed methods with HC2 and HC3 (M6-M7) have the best overall coverage across all scenarios, among all methods with covariate adjustment (M1-M7). Proposed (HC3) (M7) is slightly more conservative than Proposed (HC2) (M6). Unadjusted methods (M8-M9) have a good coverage for all cases. Although their coverage probabilities are similar to those of the proposed methods, the unadjusted methods have a much larger standard error and thus a much wider confidence interval. Therefore, they are not as efficient as the proposed methods (M6-M7).

Table~\ref{tab:secoverL} provides the standard error and the coverage probability of the 95\% confidence interval for smaller samples with the total sample sizes of 30, 60, and 90. As the total sample size increases, the standard error decreases for all methods under all scenarios. Methods M1-M7 have smaller standard errors than those from M8-M9. This reflects the efficiency gain by including prognostic covariates $X_\mathrm{cat}$ and $X_\mathrm{cont}$ in M1-M7 compared to no covariate adjustment in M8-M9. These conclusions are consistent with those based on larger sample sizes from Table~\ref{tab:secoverH}. In addition, methods using HC3 (M3 and M7) have larger standard error compared to other methods with covariate adjustment.

\begin{table}[!ht]
	\centering
	\noindent\adjustbox{max width=\textwidth}{%
		\begin{tabular}{lccc|ccc}
			\hline
			& \multicolumn{3}{c}{Standard error} & \multicolumn{3}{c}{Coverage probability}\\
			\cline{2-4} \cline{5-7}
			\multicolumn{1}{r}{Total sample size:} & 30 & 60 & 90 & 30 & 60 & 90 \\
			\multicolumn{1}{r}{Randomization ratio:} & 1:1 (2:1) & 1:1 (2:1) & 1:1 (2:1) & 1:1 (2:1) & 1:1 (2:1) & 1:1 (2:1) \\\hline
			\multicolumn{1}{l|}{Method} & \multicolumn{6}{c}{Scenario 1: Moderate treatment effect} \\\hline
			M1: Delta (model)         & 0.111 (0.193) & 0.072 (0.075) & 0.059 (0.062) & 0.883 (0.844) & 0.913 (0.903) & 0.925 (0.923) \\ 
			M2: Delta (HC2)           & 0.120 (0.117) & 0.078 (0.080) & 0.062 (0.064) & 0.902 (0.850) & 0.928 (0.910) & 0.936 (0.928) \\ 
			M3: Delta (HC3)           & 0.143 (0.143) & 0.085 (0.088) & 0.065 (0.068) & 0.940 (0.897) & 0.950 (0.934) & 0.949 (0.944) \\ 
			M4: EIF                   & 0.108 (0.106) & 0.074 (0.076) & 0.061 (0.062) & 0.887 (0.849) & 0.923 (0.906) & 0.934 (0.927) \\ 
			M5: Semi-parametric       & 0.126 (0.127) & 0.078 (0.081) & 0.062 (0.065) & 0.941 (0.915) & 0.941 (0.926) & 0.944 (0.936) \\ 
			M6: Proposed (HC2)        & 0.123 (0.122) & 0.080 (0.082) & 0.063 (0.066) & 0.923 (0.888) & 0.942 (0.927) & 0.946 (0.940) \\ 
			M7: Proposed (HC3)        & 0.146 (0.148) & 0.087 (0.091) & 0.066 (0.070) & 0.954 (0.925) & 0.960 (0.947) & 0.957 (0.953) \\ 
			M8: Unadjusted (HC2)      & 0.158 (0.163) & 0.111 (0.115) & 0.091 (0.094) & 0.935 (0.908) & 0.944 (0.935) & 0.948 (0.942) \\ 
			M9: Unadjusted (HC3)      & 0.164 (0.171) & 0.113 (0.118) & 0.092 (0.095) & 0.942 (0.918) & 0.947 (0.939) & 0.950 (0.945) \\ \hline
			\multicolumn{1}{l|}{Method} & \multicolumn{6}{c}{Scenario 2: Large treatment effect} \\\hline
			M1: Delta (model)         & 0.110 (0.168) & 0.063 (0.064) & 0.049 (0.050) & 0.895 (0.870) & 0.906 (0.885) & 0.912 (0.903) \\ 
			M2: Delta (HC2)           & 0.110 (0.103) & 0.068 (0.066) & 0.052 (0.052) & 0.903 (0.870) & 0.923 (0.884) & 0.927 (0.907) \\ 
			M3: Delta (HC3)           & 0.131 (0.127) & 0.077 (0.076) & 0.057 (0.058) & 0.936 (0.912) & 0.949 (0.915) & 0.947 (0.930) \\ 
			M4: EIF                   & 0.102 (0.097) & 0.066 (0.064) & 0.052 (0.052) & 0.894 (0.879) & 0.920 (0.892) & 0.931 (0.915) \\ 
			M5: Semi-parametric        & 0.119 (0.115) & 0.070 (0.069) & 0.054 (0.054) & 0.940 (0.933) & 0.938 (0.917) & 0.939 (0.926) \\ 
			M6: Proposed (HC2)        & 0.115 (0.109) & 0.072 (0.070) & 0.055 (0.056) & 0.922 (0.911) & 0.942 (0.916) & 0.945 (0.932) \\ 
			M7: Proposed (HC3)        & 0.134 (0.132) & 0.080 (0.079) & 0.059 (0.061) & 0.947 (0.940) & 0.961 (0.939) & 0.960 (0.949) \\ 
			M8: Unadjusted (HC2)      & 0.139 (0.140) & 0.098 (0.100) & 0.080 (0.082) & 0.933 (0.926) & 0.944 (0.932) & 0.947 (0.942) \\ 
			M9: Unadjusted (HC3)      & 0.144 (0.146) & 0.100 (0.102) & 0.081 (0.083) & 0.942 (0.931) & 0.948 (0.938) & 0.949 (0.946) \\  \hline
	\end{tabular}}
	\caption{Standard error and coverage probability of the 95\% confidence interval under 1:1 (2:1) stratified randomization and correct model specification for the total sample size of 30, 60, and 90.}
	\label{tab:secoverL}
\end{table}

In terms of the coverage probability of the 95\% confidence interval for the total sample sizes of 30, 60, and 90, methods estimating the conditional variance (M1-M3) show undercoverage for both Scenarios 1 and 2. Replacing the model-based variance (M1) with the sandwich estimators HC2 (M2) and HC3 (M3) enhances the coverage. The EIF method (M4) also displays undercoverage for both Scenarios 1 and 2. The semi-parametric approach (M5) shows a reasonable coverage when the randomization ratio is 1:1 but undercoverage when the randomization ratio is 2:1. The proposed method with HC2 (M6) also shows undercoverage. The proposed method with HC3 (M7) demonstrates more stable performance. It achieves coverage closest to 95\% under 2:1 randomization and slightly over under 1:1 randomization. The proposed unadjusted methods (M8-M9) have a reasonable coverage for all cases, except for 2:1 randomization where some level of undercoverage is evident.

In addition to standard error and the coverage probability, we also investigate the performance under Scenario 3 in Table \ref{tab:scenarios} for the null hypothesis of no treatment effect. Table~\ref{tab:alpha} shows the probability to reject the null hypothesis (or the Type I error) at the two-sided significance level of 0.05. The method estimating the conditional variance using the model-based variance (M1) shows an inflated Type I error, especially when the sample size is small. Substituting the sandwich estimators of HC2 (M2) and HC3 (M3) for the model-based variance (M1) improves the control of the Type I error. While M2 starts showing inflation with moderate sample sizes, M3 exhibits this behavior with smaller sample sizes. The method using EIF (M4) also shows an inflation of the Type I error for moderate to small sample sizes. For the semi-parametric approach (M5),  the Type I error is reasonably controlled when the randomization ratio is 1:1, but inflation occurs when the randomization ratio is 2:1, particularly with moderate to small sample sizes. The proposed method with HC2 (M6) has a similar performance as M5, except for the sample size of 30, where M6 has more inflation than M5. The proposed method with HC3 (M7) has the lowest Type I error among all methods, and sometimes is conservative when the sample size is moderate to small. The unadjusted methods have a reasonable control of the Type I error for all cases, except for 2:1 randomization with small sample sizes.

\begin{table}[!ht]
	\centering
	\noindent\adjustbox{max width=\textwidth}{%
		\begin{tabular}{lcccccc}
			\hline
			& \multicolumn{6}{c}{Type I error rate} \\\hline
			\multicolumn{1}{r}{Total sample size:} & 30 & 60 & 90 & 150 & 360 & 900 \\
			\multicolumn{1}{r}{Randomization ratio:} & 1:1 (2:1) & 1:1 (2:1) & 1:1 (2:1) & 1:1 (2:1) & 1:1 (2:1) & 1:1 (2:1)  \\\hline
			\multicolumn{1}{l|}{Method}& \multicolumn{6}{c}{Scenario 3: No treatment effect} \\\hline
			M1: Delta (model)         & 0.126 (0.170) & 0.089 (0.099) & 0.076 (0.080) & 0.064 (0.065) & 0.054 (0.055) & 0.052 (0.053) \\ 
			M2: Delta (HC2)           & 0.097 (0.155) & 0.070 (0.091) & 0.063 (0.072) & 0.057 (0.060) & 0.051 (0.054) & 0.051 (0.051) \\ 
			M3: Delta (HC3)           & 0.062 (0.114) & 0.048 (0.064) & 0.048 (0.054) & 0.049 (0.052) & 0.048 (0.050) & 0.049 (0.050) \\ 
			M4: EIF                   & 0.105 (0.161) & 0.078 (0.102) & 0.067 (0.079) & 0.059 (0.064) & 0.052 (0.055) & 0.051 (0.052) \\ 
			M5: Semi-parametric       & 0.052 (0.092) & 0.056 (0.078) & 0.056 (0.069) & 0.054 (0.060) & 0.050 (0.053) & 0.050 (0.052) \\ 
			M6: Proposed (HC2)        & 0.072 (0.124) & 0.056 (0.077) & 0.054 (0.063) & 0.052 (0.056) & 0.049 (0.052) & 0.050 (0.051) \\ 
			M7: Proposed (HC3)        & 0.046 (0.088) & 0.038 (0.053) & 0.041 (0.047) & 0.045 (0.047) & 0.047 (0.048) & 0.049 (0.050) \\ 
			M8: Unadjusted (HC2)      & 0.061 (0.103) & 0.056 (0.067) & 0.052 (0.059) & 0.051 (0.054) & 0.050 (0.051) & 0.051 (0.051) \\ 
			M9: Unadjusted (HC3)      & 0.056 (0.097) & 0.050 (0.061) & 0.050 (0.056) & 0.049 (0.053) & 0.049 (0.050) & 0.051 (0.050) \\ 
			\hline
	\end{tabular}}
	\caption{Type I error rate at the two-sided significance level of 0.05 under 1:1 (2:1) stratified randomization.}
	\label{tab:alpha}
\end{table}

To summarize the performance based on Tables~\ref{tab:secoverH}, \ref{tab:secoverL}, and \ref{tab:alpha}, we observe different behaviors for large samples and small samples. When the sample size is large, the proposed variance estimator with HC2 (M6) appears to have the best performance with the 95\% confidence interval coverage closest to 95\% under both 1:1 and 2:1 randomization ratios. In contrast, when dealing with small sample sizes, the proposed variance estimator with HC3 (M7) has the best overall performance both 1:1 and 2:1 randomization. When the randomization ratio is 1:1, the semi-parametric approach (M5) displays reasonable performance across small to large sample sizes, although its performance is affected by the 2:1 randomization ratio with small sample sizes. These conclusions also hold for the simple randomization, whose results are provided in the Supplementary Materials. Note that these conclusions are dependent on the true parameters in Table~\ref{tab:scenarios}. For more rare or more frequent outcomes, the conclusions may change and additional simulations may be needed.

\subsection{Model misspecification}\label{sec:simmiss}
To investigate model misspecification, we generate the binary response $Y$ from the following logistic regression model:
$$
\mathrm{logit}\left\{P(Y=1|Z, X_\mathrm{cont}, X_\mathrm{cat})\right\}=\beta_{0}+\beta_{1}Z + \beta_{2} X_\mathrm{cont} + \beta_{3} X_\mathrm{cat} + \beta_{4}X_\mathrm{cont}^2 + \beta_{5} X_\mathrm{cont}\times Z  + \beta_{6} X_\mathrm{cont}^2\times Z,
$$
where true parameters are listed in Table~\ref{tab:misscenarios}. In scenario 4, $Y$ is dependent on $X_\mathrm{cont}\times Z$, $X_\mathrm{cont}^2$ and $X_\mathrm{cont}^2\times Z$, all of which are neglected in standardization \eqref{eq:logistic}. In scenario 5, $Y$ relies only on $Z$ ($\beta_{2} = \cdots =\beta_{6} = 0)$, indicating that standardization  \eqref{eq:logistic} includes more covariates than necessary.

\begin{table}[!ht]
	\centering
	\begin{tabular}{ccccc}
		\hline
		Scenario & $(\beta_{0},\beta_{1},\beta_{2},\beta_{3},\beta_{4},\beta_{5},\beta_{6})$  & $\pi_{(0)}$ & $\pi_{(1)}$ & RD \\ \hline
		4 &$( - 4,   2,  4.2,  -3,   1,  -0.2, 0.2)$ & $0.17$ & $0.28$ & $0.11$ \\ 
		5 &$(-2.2, 0.7,    0,   0,   0,   0,    0 )$ & $0.10$ & $0.18$ & $0.08$ \\ 
		\hline
	\end{tabular}
	\caption{True parameters for simulation under model misspecification.} \label{tab:misscenarios}
\end{table}

Tables~\ref{tab:secoverMH} and \ref{tab:secoverML} present the standard error and the coverage probability of the 95\% confidence interval. Conclusions are consistent with those made under the correct model specification. Specifically, for scenarios with large sample sizes, the proposed variance estimator with HC2 (M6) seems to perform the best, achieving a 95\% confidence interval coverage closest to 95\% under both 1:1 and 2:1 randomization ratios. In scenarios with small sample sizes, the proposed variance estimator with HC3 (M7) has the most favorable overall performance for both 1:1 and 2:1 randomization ratios. For the 1:1 randomization ratio, the semi-parametric approach (M5) shows reasonable performance across a wide range of sample sizes, though its performance is impacted when dealing with the 2:1 randomization ratio with small sample sizes. Notably, in Scenario 5, where the data generating model excludes covariates, the unadjusted methods (M8-M9) should theoretically perform better due to their correct model specification. However, their 95\% confidence interval coverage falls below 95\% for small sample sizes, indicating the challenges of small samples for all methods.

\begin{table}[!ht]
	\centering
	\noindent\adjustbox{max width=\textwidth}{%
		\begin{tabular}{lccc|ccc}
			\hline
			& \multicolumn{3}{c}{Standard error} & \multicolumn{3}{c}{Coverage probability}\\
			\cline{2-4} \cline{5-7}
			\multicolumn{1}{r}{Total sample size:} & 150 & 360 & 900 & 150 & 360 & 900 \\
			\multicolumn{1}{r}{Randomization ratio:} & 1:1 (2:1) & 1:1 (2:1) & 1:1 (2:1) & 1:1 (2:1) & 1:1 (2:1) & 1:1 (2:1) \\\hline
			\multicolumn{1}{l|}{Method}& \multicolumn{6}{c}{Scenario 4: Model misspecification with missing covariates} \\\hline
			M1: Delta (model)         & 0.037 (0.039) & 0.024 (0.025) & 0.015 (0.016) & 0.926 (0.931) & 0.935 (0.940) & 0.938 (0.943) \\ 
			M2: Delta (HC2)           & 0.039 (0.039) & 0.024 (0.025) & 0.015 (0.015) & 0.935 (0.928) & 0.936 (0.935) & 0.938 (0.937) \\ 
			M3: Delta (HC3)           & 0.040 (0.041) & 0.024 (0.025) & 0.015 (0.016) & 0.945 (0.941) & 0.941 (0.940) & 0.939 (0.939) \\ 
			M4: EIF                   & 0.039 (0.040) & 0.025 (0.026) & 0.016 (0.016) & 0.942 (0.937) & 0.948 (0.946) & 0.950 (0.949) \\ 
			M5: Semi-parametric       & 0.040 (0.040) & 0.025 (0.026) & 0.016 (0.016) & 0.944 (0.939) & 0.949 (0.946) & 0.950 (0.949) \\ 
			M6: Proposed (HC2)        & 0.041 (0.042) & 0.025 (0.026) & 0.016 (0.016) & 0.950 (0.947) & 0.951 (0.950) & 0.951 (0.950) \\ 
			M7: Proposed (HC3)        & 0.042 (0.043) & 0.026 (0.026) & 0.016 (0.016) & 0.958 (0.957) & 0.954 (0.954) & 0.953 (0.952) \\ 
			M8: Unadjusted (HC2)      & 0.068 (0.069) & 0.044 (0.045) & 0.028 (0.028) & 0.948 (0.945) & 0.950 (0.950) & 0.950 (0.950) \\ 
			M9: Unadjusted (HC3)      & 0.068 (0.070) & 0.044 (0.045) & 0.028 (0.028) & 0.950 (0.947) & 0.951 (0.951) & 0.951 (0.950) \\ \hline
			\multicolumn{1}{l|}{Method} & \multicolumn{6}{c}{Scenario 5: Model misspecification with additional unnecessary covariates} \\\hline
			M1: Delta (model)         & 0.056 (0.057) & 0.036 (0.037) & 0.023 (0.023) & 0.945 (0.942) & 0.948 (0.948) & 0.948 (0.948) \\ 
			M2: Delta (HC2)           & 0.057 (0.058) & 0.037 (0.037) & 0.023 (0.023) & 0.949 (0.946) & 0.949 (0.949) & 0.949 (0.949) \\ 
			M3: Delta (HC3)           & 0.058 (0.058) & 0.037 (0.037) & 0.023 (0.023) & 0.952 (0.949) & 0.951 (0.950) & 0.949 (0.950) \\ 
			M4: EIF                   & 0.056 (0.057) & 0.036 (0.037) & 0.023 (0.023) & 0.946 (0.943) & 0.948 (0.948) & 0.948 (0.948) \\ 
			M5: Semi-parametric       & 0.058 (0.058) & 0.037 (0.037) & 0.023 (0.023) & 0.952 (0.949) & 0.951 (0.950) & 0.949 (0.950) \\ 
			M6: Proposed (HC2)        & 0.057 (0.058) & 0.037 (0.037) & 0.023 (0.023) & 0.949 (0.946) & 0.949 (0.949) & 0.949 (0.949) \\ 
			M7: Proposed (HC3)        & 0.058 (0.059) & 0.037 (0.037) & 0.023 (0.023) & 0.952 (0.949) & 0.951 (0.950) & 0.949 (0.950) \\ 
			M8: Unadjusted (HC2)      & 0.056 (0.057) & 0.036 (0.037) & 0.023 (0.023) & 0.948 (0.946) & 0.950 (0.949) & 0.949 (0.949) \\ 
			M9: Unadjusted (HC3)      & 0.057 (0.058) & 0.037 (0.037) & 0.023 (0.023) & 0.950 (0.948) & 0.950 (0.950) & 0.949 (0.949) \\ 
			\hline
	\end{tabular}}
	\caption{Standard error and coverage probability of the 95\% confidence interval under 1:1 (2:1) stratified randomization and model misspecification for the total sample size of 150, 360 and 900.}
	\label{tab:secoverMH}
\end{table}

\begin{table}[!ht]
	\centering
	\noindent\adjustbox{max width=\textwidth}{%
		\begin{tabular}{lccc|ccc}
			\hline
			& \multicolumn{3}{c}{Standard error} & \multicolumn{3}{c}{Coverage probability}\\
			\cline{2-4} \cline{5-7}
			\multicolumn{1}{r}{Total sample size:} & 30 & 60 & 90 & 30 & 60 & 90 \\
			\multicolumn{1}{r}{Randomization ratio:} & 1:1 (2:1) & 1:1 (2:1) & 1:1 (2:1) & 1:1 (2:1) & 1:1 (2:1) & 1:1 (2:1) \\\hline
			\multicolumn{1}{l|}{Method} & \multicolumn{6}{c}{Scenario 4: Model misspecification with missing covariates} \\\hline
			M1: Delta (model)         & 0.119 (0.185) & 0.065 (0.068) & 0.050 (0.052) & 0.909 (0.876) & 0.918 (0.906) & 0.919 (0.921) \\ 
			M2: Delta (HC2)           & 0.117 (0.110) & 0.070 (0.068) & 0.053 (0.053) & 0.911 (0.862) & 0.929 (0.894) & 0.931 (0.916) \\ 
			M3: Delta (HC3)           & 0.136 (0.134) & 0.078 (0.077) & 0.057 (0.058) & 0.943 (0.904) & 0.953 (0.925) & 0.950 (0.938) \\ 
			M4: EIF                   & 0.109 (0.104) & 0.068 (0.065) & 0.053 (0.052) & 0.901 (0.867) & 0.927 (0.898) & 0.934 (0.922) \\ 
			M5: Semi-parametric       & 0.126 (0.122) & 0.071 (0.070) & 0.054 (0.054) & 0.947 (0.922) & 0.942 (0.919) & 0.940 (0.931) \\ 
			M6: Proposed (HC2)        & 0.121 (0.115) & 0.073 (0.072) & 0.056 (0.056) & 0.928 (0.899) & 0.945 (0.921) & 0.947 (0.938) \\ 
			M7: Proposed (HC3)        & 0.140 (0.139) & 0.081 (0.081) & 0.060 (0.061) & 0.952 (0.931) & 0.963 (0.943) & 0.961 (0.955) \\ 
			M8: Unadjusted (HC2)      & 0.152 (0.155) & 0.107 (0.110) & 0.087 (0.090) & 0.937 (0.915) & 0.945 (0.935) & 0.948 (0.943) \\ 
			M9: Unadjusted (HC3)      & 0.158 (0.162) & 0.109 (0.112) & 0.088 (0.091) & 0.942 (0.924) & 0.949 (0.941) & 0.951 (0.946) \\ \hline 
			\multicolumn{1}{l|}{Method} & \multicolumn{6}{c}{Scenario 5: Model misspecification with additional unnecessary covariates} \\\hline
			M1: Delta (model)         & 0.127 (0.359) & 0.087 (0.088) & 0.072 (0.073) & 0.885 (0.898) & 0.933 (0.921) & 0.940 (0.932) \\ 
			M2: Delta (HC2)           & 0.131 (0.131) & 0.091 (0.092) & 0.074 (0.075) & 0.907 (0.926) & 0.942 (0.932) & 0.946 (0.939) \\ 
			M3: Delta (HC3)           & 0.148 (0.147) & 0.095 (0.096) & 0.076 (0.077) & 0.933 (0.951) & 0.953 (0.944) & 0.952 (0.945) \\ 
			M4: EIF                   & 0.121 (0.121) & 0.088 (0.089) & 0.072 (0.073) & 0.889 (0.908) & 0.934 (0.925) & 0.941 (0.935) \\ 
			M5: Semi-parametric       & 0.138 (0.138) & 0.094 (0.095) & 0.075 (0.076) & 0.919 (0.945) & 0.949 (0.941) & 0.951 (0.944) \\ 
			M6: Proposed (HC2)        & 0.132 (0.132) & 0.091 (0.092) & 0.074 (0.075) & 0.912 (0.934) & 0.944 (0.935) & 0.947 (0.940) \\ 
			M7: Proposed (HC3)        & 0.149 (0.148) & 0.095 (0.096) & 0.076 (0.077) & 0.937 (0.957) & 0.955 (0.947) & 0.953 (0.947) \\ 
			M8: Unadjusted (HC2)      & 0.125 (0.125) & 0.089 (0.090) & 0.073 (0.074) & 0.912 (0.930) & 0.943 (0.933) & 0.946 (0.938) \\ 
			M9: Unadjusted (HC3)      & 0.130 (0.131) & 0.091 (0.092) & 0.074 (0.075) & 0.918 (0.935) & 0.948 (0.938) & 0.949 (0.942) \\ 
			\hline
	\end{tabular}}
	\caption{Standard error and coverage probability of the 95\% confidence interval under 1:1 (2:1) stratified randomization and model misspecification for the total sample size of 30, 60, and 90.}
	\label{tab:secoverML}
\end{table}

\section{Case Study} \label{sec:case}
We illustrate our methods using data from the iron deficiency study in Peru on reducing anemia among adolescents by a low-cost encouragement intervention \cite{chong2016iron}. Code examples are provided to illustrate the simple implementation of our proposal. Participants were randomly exposed to a ``placebo" video featured a dentist promoting oral hygiene without mentioning iron at all, or one of two ``treatment" videos encouraging iron supplements under equal randomization. The first ``treatment" (Soccer) shows a popular soccer player encouraging iron supplements to maximize energy, and the second (Physician) shows a doctor encouraging iron supplements for overall health. In total, there are 72, 70 and 73 participants in the placebo group and two treatment groups, respectively. The dataset is available at \url{https://www.openicpsr.org/openicpsr/project/113624/version/V1/view}. 

The outcome of interest is a binary variable indicating whether a student was anemic, determined through hemoglobin tests measured during the follow-up survey. We use the same set of baseline covariates as in the original analyses \cite{chong2016iron}, which includes the student's gender, monthly income, availability of electricity at home, and mother's years of schooling. Following a similar approach \cite{chong2016iron}, we conducted separate analyses for participants who suffered from iron deficiency anemia (IDA) at baseline and who were not anemic. For brevity, we present the analysis solely for those who suffered from IDA at baseline. The respective groups for the placebo, Soccer and Physician interventions comprise 29, 27 and 32 participants, respectively.

First, we fit a logistic regression to adjust for baseline covariates. In this case, the regression coefficient of the treatment assignment is a conditional estimator, which estimates the change in the log odds of the outcome with a change of treatment from placebo to one of the two treatments when holding the baseline outcome constant. The treatment effect is estimated to be an odds ratio of 0.179 (0.027, 1.183) with a $p$-value of $0.074$ for the Physician group and 0.323 (0.064, 1.629) with a $p$-value of $0.171$ for the Soccer group, where the 95\% confidence interval is derived using the sandwich estimator HC3.

Second, we focus on the unconditional treatment effect averaging over the entire population. To achieve it, we apply the standardization on top of the logistic regression model using the proposed sandwich variance HC3 \eqref{eq:impvar}. In this case, we report the risk difference, which is easier to communicate than the odds ratio. The treatment effect for the Physician group is estimated to be a risk difference of -0.274 (-0.529, -0.019) with a $p$-value of $0.035$, and for the Soccer group, it is -0.199 (-0.489, 0.092) with a $p$-value of $0.18$. In addition, we report the unadjusted risk difference using only the treatment assignment in the logistic regression model. For the Physician group, the unadjusted risk difference is -0.211 (-0.465, 0.043) with a $p$-value of $0.103$, and for the Soccer group, it is -0.179 (-0.446, 0.089) with a $p$-value of $0.19$. The Soccer group shows no positive treatment effect on the anemia rate in both the adjusted and unadjusted analysis. The Physician group, on the other hand, demonstrates a positive treatment effect on the anemia rate after adjusting for baseline covariates. For both comparisons, we can see that adjusting for covariates improve the efficiency of the standardization estimator because of narrower confidence intervals.

One advantage of our proposed method lies in implementation by using existing R packages {\tt margins} and {\tt sandwich}. Here we illustrate the steps using pseudocode examples below.

\begin{lstlisting}[language=R]
# Load packages 
library(sandwich)
library(margins)

# Logistic regression
logistic_regression <- glm(formula = y ~ trt + covariate1 + covariate2,
                           data = data,
                           family = binomial
)

# Unconditional treatment effect using "margins" and "sandwich" packages
treatment_effect <- margins::margins(model = logistic_regression,
                                     variables = "trt",
                                     vcov = vcovHC(logistic_regression, 
                                                   type = "HC3") # Or "HC2"
)

# Summary of treatment_effect
summary_effect <- summary(treatment_effect)

# unconditional treatment effect
est <- summary_effect$AME

# Unconditional standard error
# Note "dydx_trt" depends on the variable name "trt" for the treatment assignment
se <- sqrt(summary_effect$SE^2 + var(treatment_effect$dydx_trt)/nrow(data))

# Print results
round(data.frame(estimate = est,
                 lower = est - se * qnorm(1 - 0.025),
                 upper = est + se * qnorm(1 - 0.025),
                 pvalue = 2 * (1 - pnorm(abs(est / se)))), 3)
\end{lstlisting}

\section{Discussion} \label{sec:conclusion}

We explore the use of standardization in estimating unconditional differences in proportions in randomized clinical trials. It has been recognized that the standardization estimator improves efficiency by adjusting for baseline covariates while maintaining robustness against model misspecification \cite{vanlancker2022combining}. However, ensuring valid inference under model misspecification and small sample sizes is equally important for practitioners. Our findings indicate that the conditional variance estimator based on the delta method \citep{ge2011covariate} tends to underestimate the unconditional variances even with large samples, prompting us to propose an unconditional variance estimator. 

In addition, to ensure robust performances with small sample sizes, we adopt the Huber-White robust ``sandwich” estimator. Our extensive simulations demonstrate the robustness of our proposed variance estimator across various sample sizes, randomization ratios, and both correctly and incorrectly specified models. For completeness, our method is compared to various alternative variance estimation approaches, such as the efficient influence functions, and a more recently proposed semi-parametric approach \cite{ye2023robust}, showcasing comparable or superior performances. 

In summary, our proposed method, coupled with the robust variance estimator, shows promise for wider integration into clinical trial practice owing to its consistent and robust performance. An added benefit is its compatibility with existing R packages, thereby enhancing its practicality for practitioners. In the context of large sample sizes, we recommend adopting of the proposed variance estimator with HC2. In the instances of smaller sample sizes, there exists a potential for the underestimation of variance. To address this limitation, we advocate for the preferential use of HC3 over HC2, as HC3 offers improved coverage. The semi-parametric approach also demonstrates reasonable performance across a wide spectrum of scenarios \cite{ye2023robust}. Yet, its performance could be compromised when the sample size is limited. We note that the bootstrap could also be considered; however, it is time consuming and prone to convergence problems in small sample sizes. 

Recent research allows one to replace the logistic regression model with more flexible alternatives, incorporating covariate interactions, higher-order terms, or even utilizing machine learning methods, all while ensuring valid inferences \cite{tsiatis2008covariate, chernozhukov2017double}. These approaches have the potential to enhance efficiency, albeit typically requiring larger sample sizes. In this article, our focus lies in exploring smaller sample sizes, which often makes it feasible only to consider simpler models.

As a final note, although our focus in this work is on differences in proportions, our proposed unconditional variance estimator can be readily extended to other summary measures, such as the ratio of proportions and odds ratio. These extensions will be left for future research endeavors. 

\bibliographystyle{plainnat}
\bibliography{covadj}

\appendix
\section{Review of other variance estimation methods} \label{sec:variance}
\subsection{Robust variance estimation based on semi-parametric approaches} \label{sec:ye}
A variance estimator based on a doubly robust representation of the standardization estimator is as follows \cite{ye2023robust}: 
$$
\widehat{\pi}_{(j)} = \frac{1}{n}\sum_{i=1}^n  \widehat{\pi}_{i(j)} + \frac{1}{n}\sum_{i=1}^n\left\{\left(\frac{Z_i}{n_1/n}+\frac{1-Z_i}{1-n_1/n}\right)\left( Y^{(j)}_i 
- \widehat{\pi}_{i(j)} \right)  \right\},
$$
where $n_1$ is the number of subjects in the treatment group, and $Y^{(j)}_i $ is the potential outcome of subject $i$ were they assigned to treatment $j$. Note that the first term is the standardization estimator itself, while the second term is 0 based on the first-order conditions of the maximum likelihood estimation of the logistic regression. After rewriting this equation and leveraging the results from semi-parametric approaches \cite{kennedy2016semiparametric, chernozhukov2017double}, the asymptotic distribution of the standardization estimator is
$$
\widehat{\pi}_{(j)} = \frac{1}{n}\sum_{i=1}^n \left\{\left(\frac{Z_i}{\theta}+\frac{1-Z_i}{1-\theta}\right)\left( Y^{(j)}_i 
- \pi_{i(j)}^{*} \right) + \pi_{i(j)}^{*} \right\} + o_p\left(n^{-1/2}\right),
$$
where $\theta = \Pr(Z=1)$, $\pi_{i(j)}^{*}$ is the probability limit of $\widehat{\pi}_{i(j)}$ when $n$ goes to infinity, and $o_p\left(n^{-1/2}\right)$ denotes remaining terms divided by $n^{-1/2}$ converges to 0 in probability as $n$ goes to infinity.

Applying the central limit theorem, the asymptotic variance of $\widehat{\pi}_{(j)}$ and covariance between $\widehat{\pi}_{(0)}$ and $\widehat{\pi}_{(1)}$ are given by:
$$
\Var\left(\widehat{\pi}_{(j)}\right) = n^{-1}\left\{\theta^{-1} \Var\left( Y^{(j)} - \pi_{(j)}^{*} \right) + 2\Cov\left( Y^{(j)}, \pi_{(j)}^{*} \right) - \Var\left(\pi_{(j)}^{*}\right)\right\} , j = 0,1,
$$ 
$$
\Cov\left(\widehat{\pi}_{(0)}, \widehat{\pi}_{(1)}\right) = n^{-1}\left\{\Cov\left( Y^{(0)}, \pi_{(1)}^{*} \right) + \Cov\left( Y^{(1)}, \pi_{(0)}^{*} \right) - \Cov\left(\pi_{(0)}^{*},\pi_{(1)}^{*}\right)\right\}.
$$ 
Here, $\Var\left( Y^{(j)} - \pi_{(j)}^{*} \right)$ is estimated by the sample variance of $Y_i - \widehat{\pi}_{i(j)}$, while $\Cov\left( Y^{(j)}, \pi_{(k)}^{*} \right), k = 0,1$ is estimated by the sample covariance of $Y_i$ and $\widehat{\pi}_{i(k)}$, using subjects in treatment $j$. Furthermore, $\Var\left(\pi_{(j)}^{*}\right)$ is estimated by the sample variance of $\widehat{\pi}_{i(j)}$, while $ \Cov\left( Y^{(0)}, \pi_{(1)}^{*} \right)$ is estimated by the sample covariance of $\widehat{\pi}_{i(0)}$ and $\widehat{\pi}_{i(1)}$ using all subjects in both treatment groups. By the delta method, 
$$
\Var\left(\widehat{\mathrm{RD}}\right) = \Var\left(\widehat{\pi}_{(1)}\right) - 2\Cov\left(\widehat{\pi}_{(0)}, \widehat{\pi}_{(1)}\right)+ \Var\left(\widehat{\pi}_{(0)}\right) .
$$
This approach is implemented in the R package {\tt RobinCar} \cite{robincar2022}. 

\subsection{Efficient influence function approach} \label{sec:eif}
The efficient influence function is a popular approach in targeted learning \citep{van2011targeted}, because they effectively enable the use of data-adaptive estimation strategies to model the data-generating distribution, whilst permitting valid inference of the estimand of interest \citep{hines2022demystifying}. The efficient influence function for the logistic regression \eqref{eq:logistic} \citep{jann2023estimation} is defined as 
$$
\bm{\lambda}_i (\bm{b}) = \left\{ \frac{1}{n} \sum_{i=1}^n \widehat{\pi}_i\left(1-\widehat{\pi}_i\right) \bm{x}_i^{\top}\bm{x}_i\right\}^{-1}\bm{x}_i\left(y_i-\widehat{\pi}_i\right),
$$
which represents the sensitivity of $\bm{b}$ to perturbations in the observed data for subject $i$. By the delta method, the efficient influence function for $\widehat{\pi}_{(j)}, j=0, 1$ is
$$
\bm{\lambda}_{i}\left(\widehat{\pi}_{(j)}\right)=\left(\widehat{\pi}_{i(j)} - \widehat{\pi}_{(j)}\right) +  \left(\frac{\partial \widehat{\pi}_{(j)}}{\partial \bm{b}} \right)^\top \lambda_i (\bm{b})=\left(\widehat{\pi}_{i(j)} - \widehat{\pi}_{(j)}\right) + \bm{d}_{(j)}^\top\lambda_i (\bm{b}),
$$
where $\bm{d}_{(j)} = \frac{1}{n}\sum_{i=1}^{n} \bm{x}_{i(j)}\widehat{\pi}_{i(j)} \left(1-\widehat{\pi}_{i(j)}\right).$

Thus the efficient influence function of RD is:
$$
\bm{\lambda}_i\left(\widehat{\mathrm{RD}}\right) =  \bm{\lambda}_i\left(\widehat{\pi}_{(1)}\right) - \bm{\lambda}_i\left(\widehat{\pi}_{(0)}\right).
$$
The variance of $\widehat{\mathrm{RD}}$ can be estimated by $1/n$ times the sample variance of the efficient influence function on the right hand side. The confidence interval for $\widehat{\mathrm{RD}}$ is given by \eqref{eq:ci}.

\section{Simulation results for other versions of the sandwich estimator} 
\label{sec:allhc}

In Tables \ref{tab:secoverallH}--\ref{tab:secoverallML}, we provide simulation results for all types of sandwich variance-covariance matrices included in the {\tt sandwich} R package \citep{zeileis2020various} as well as the proposed method using the model-based variance estimator (Proposed (model)). Tables \ref{tab:secoverallH}--\ref{tab:secoverallL} show the standard error and the coverage probability of the 95\% confidence interval under correct model specification. When dealing with larger sample sizes 150, 360, and 900, Proposed (model), Proposed (HC0) and Proposed (HC1) show slight undercoverage for Scenario 1 and a more pronounced undercoverage for Scenario 2. On the contrary, Proposed (const), Proposed (HC4) and Proposed (HC4m) show overcoverage. Proposed (HC2) seems to have the best performance with the 95\% confidence interval coverage closest to the nominal level under both randomization ratios, followed by Proposed (HC3) and Proposed (HC5). When dealing with small sample sizes 30, 60, and 90, Proposed (model), Proposed (HC0) and Proposed (HC1) fail to provide enough coverage, while Proposed (HC4) and Proposed (HC4m) show slight overcoverage. Proposed (const) tends to yield unreasonably wide confidence intervals when per-group sample sizes are small. Proposed (HC3) has the best overall performance both 1:1 and 2:1 randomization. The conclusions are similar with those made under the incorrect model specification, as shown in Tables \ref{tab:secoverallMH}--\ref{tab:secoverallML}.

In terms of the performance under Scenario 3 for the null hypothesis of no treatment effect, we present the probability to reject the null hypothesis (or the Type I error) at the two-sided significance level of 0.05 in Table~\ref{tab:alphaall}. Proposed (model), Proposed (HC0) and Proposed (HC1) show inflated Type I errors, especially when the sample size is small. In contrast, Proposed (HC4) and Proposed (HC4m) are conservative even with moderate sample sizes. Proposed (HC2) and Proposed (HC5) have the Type I error reasonably controlled when sample size is moderate to large but show slight inflation of the Type I error when sample sizes are small. Proposed (HC3) has the lowest Type I error among all methods, and sometimes is conservative when the sample size is moderate to small.

\begin{table}[!ht]
	\centering
	\noindent\adjustbox{max width=\textwidth}{%
		\begin{tabular}{lccc|ccc}
			\hline
			& \multicolumn{3}{c}{Standard error} & \multicolumn{3}{c}{Coverage probability}\\ \cline{2-4} \cline{5-7}
			\multicolumn{1}{r}{Total sample size:} & 150 & 360 & 900 & 150 & 360 & 900 \\
			\multicolumn{1}{r}{Randomization ratio:} & 1:1 (2:1) & 1:1 (2:1) & 1:1 (2:1) & 1:1 (2:1) & 1:1 (2:1) & 1:1 (2:1) \\\hline
			\multicolumn{1}{l|}{Method} & \multicolumn{6}{c}{Scenario 1: Moderate treatment effect} \\\hline
			Proposed (model) & 0.047 (0.049) & 0.030 (0.032) & 0.019 (0.020) & 0.942 (0.942) & 0.948 (0.948) & 0.950 (0.950) \\ 
			Proposed (const) & 0.049 (0.053) & 0.031 (0.033) & 0.019 (0.021) & 0.952 (0.959) & 0.952 (0.959) & 0.952 (0.959) \\ 
			Proposed (HC0)   & 0.047 (0.049) & 0.030 (0.032) & 0.019 (0.020) & 0.942 (0.938) & 0.948 (0.947) & 0.950 (0.950) \\ 
			Proposed (HC1)   & 0.048 (0.049) & 0.031 (0.032) & 0.019 (0.020) & 0.945 (0.942) & 0.949 (0.948) & 0.951 (0.950) \\ 
			Proposed (HC2)   & 0.048 (0.050) & 0.031 (0.032) & 0.019 (0.020) & 0.948 (0.946) & 0.950 (0.950) & 0.951 (0.951) \\ 
			Proposed (HC3)   & 0.050 (0.052) & 0.031 (0.032) & 0.019 (0.020) & 0.954 (0.953) & 0.953 (0.953) & 0.952 (0.952) \\ 
			Proposed (HC4)   & 0.050 (0.053) & 0.031 (0.033) & 0.019 (0.020) & 0.957 (0.960) & 0.953 (0.955) & 0.952 (0.953) \\ 
			Proposed (HC4m)  & 0.050 (0.053) & 0.031 (0.033) & 0.019 (0.020) & 0.957 (0.957) & 0.954 (0.954) & 0.952 (0.953) \\ 
			Proposed (HC5)   & 0.048 (0.051) & 0.031 (0.032) & 0.019 (0.020) & 0.950 (0.950) & 0.951 (0.951) & 0.951 (0.951) \\ 
			\hline
			\multicolumn{1}{l|}{Method} & \multicolumn{6}{c}{Scenario 2: Large treatment effect} \\\hline
			Proposed (model) & 0.040 (0.041) & 0.026 (0.026) & 0.016 (0.017) & 0.941 (0.941) & 0.946 (0.947) & 0.948 (0.949) \\ 
			Proposed (const) & 0.043 (0.047) & 0.026 (0.029) & 0.016 (0.018) & 0.958 (0.967) & 0.952 (0.964) & 0.950 (0.963) \\ 
			Proposed (HC0)   & 0.040 (0.040) & 0.026 (0.026) & 0.016 (0.017) & 0.939 (0.934) & 0.946 (0.945) & 0.948 (0.948) \\ 
			Proposed (HC1)   & 0.040 (0.041) & 0.026 (0.026) & 0.016 (0.017) & 0.942 (0.937) & 0.947 (0.946) & 0.949 (0.948) \\ 
			Proposed (HC2)   & 0.041 (0.042) & 0.026 (0.027) & 0.016 (0.017) & 0.949 (0.945) & 0.949 (0.949) & 0.949 (0.949) \\ 
			Proposed (HC3)   & 0.043 (0.044) & 0.026 (0.027) & 0.016 (0.017) & 0.957 (0.955) & 0.953 (0.953) & 0.951 (0.951) \\ 
			Proposed (HC4)   & 0.045 (0.048) & 0.027 (0.028) & 0.016 (0.017) & 0.969 (0.970) & 0.956 (0.958) & 0.952 (0.953) \\ 
			Proposed (HC4m)  & 0.044 (0.045) & 0.026 (0.027) & 0.016 (0.017) & 0.962 (0.960) & 0.954 (0.955) & 0.951 (0.951) \\ 
			Proposed (HC5)   & 0.043 (0.046) & 0.026 (0.027) & 0.016 (0.017) & 0.958 (0.960) & 0.952 (0.953) & 0.950 (0.951) \\ 
			\hline
	\end{tabular}}
	\caption{Standard error and coverage probability of the 95\% confidence interval under 1:1 (2:1) stratified randomization and correct model specification for the total sample size of 150, 360, and 900.}
	\label{tab:secoverallH}
\end{table}

\begin{table}[!ht]
	\centering
	\noindent\adjustbox{max width=\textwidth}{%
		\begin{tabular}{lccc|ccc}
			\hline
			& \multicolumn{3}{c}{Standard error} & \multicolumn{3}{c}{Coverage probability}\\
			\cline{2-4} \cline{5-7}
			\multicolumn{1}{r}{Total sample size:} & 30 & 60 & 90 & 30 & 60 & 90 \\
			\multicolumn{1}{r}{Randomization ratio:} & 1:1 (2:1) & 1:1 (2:1) & 1:1 (2:1) & 1:1 (2:1) & 1:1 (2:1) & 1:1 (2:1) \\\hline
			\multicolumn{1}{l|}{Method} & \multicolumn{6}{c}{Scenario 1: Moderate treatment effect} \\\hline
			Proposed (model) & 0.115 (0.198) & 0.074 (0.078) & 0.060 (0.063) & 0.910 (0.884) & 0.929 (0.922) & 0.936 (0.935) \\ 
			Proposed (const) & 3.923  (336.276) & 0.087 (0.209) & 0.066 (0.074) & 0.953 (0.952) & 0.959 (0.962) & 0.956 (0.962) \\ 
			Proposed (HC0)   & 0.107 (0.105) & 0.074 (0.075) & 0.060 (0.062) & 0.883 (0.842) & 0.921 (0.903) & 0.932 (0.925) \\ 
			Proposed (HC1)   & 0.113 (0.111) & 0.076 (0.077) & 0.062 (0.063) & 0.898 (0.861) & 0.929 (0.911) & 0.938 (0.931) \\ 
			Proposed (HC2)   & 0.123 (0.122) & 0.080 (0.082) & 0.063 (0.066) & 0.923 (0.888) & 0.942 (0.927) & 0.946 (0.940) \\ 
			Proposed (HC3)   & 0.146 (0.148) & 0.087 (0.091) & 0.066 (0.070) & 0.954 (0.925) & 0.960 (0.947) & 0.957 (0.953) \\ 
			Proposed (HC4)   & 0.196 (158.729) & 0.096 (0.106) & 0.069 (0.075) & 0.968 (0.940) & 0.974 (0.966) & 0.966 (0.967) \\ 
			Proposed (HC4m)  & 0.160 (0.194) & 0.090 (0.095) & 0.068 (0.072) & 0.963 (0.936) & 0.967 (0.955) & 0.962 (0.959) \\ 
			Proposed (HC5)   & 0.134 (9.418) & 0.084 (0.089) & 0.064 (0.069) & 0.940 (0.906) & 0.953 (0.944) & 0.952 (0.951) \\ 
			\hline
			\multicolumn{1}{l|}{Method} & \multicolumn{6}{c}{Scenario 2: Large treatment effect} \\\hline
			Proposed (model) & 0.114 (0.174) & 0.067 (0.068) & 0.052 (0.053) & 0.919 (0.916) & 0.930 (0.916) & 0.935 (0.930) \\ 
			Proposed (const) & 35.804 (249.386) & 0.089 (0.872) & 0.060 (0.093) & 0.939 (0.950) & 0.962 (0.960) & 0.962 (0.966) \\ 
			Proposed (HC0)   & 0.101 (0.096) & 0.065 (0.064) & 0.052 (0.052) & 0.890 (0.874) & 0.917 (0.889) & 0.929 (0.913) \\ 
			Proposed (HC1)   & 0.106 (0.100) & 0.067 (0.065) & 0.053 (0.053) & 0.901 (0.888) & 0.925 (0.897) & 0.934 (0.918) \\ 
			Proposed (HC2)   & 0.115 (0.109) & 0.072 (0.070) & 0.055 (0.056) & 0.922 (0.911) & 0.942 (0.916) & 0.945 (0.932) \\ 
			Proposed (HC3)   & 0.134 (0.132) & 0.080 (0.079) & 0.059 (0.061) & 0.947 (0.940) & 0.961 (0.939) & 0.960 (0.949) \\ 
			Proposed (HC4)   & 0.190 (396.900) & 0.099 (0.104) & 0.067 (0.072) & 0.955 (0.949) & 0.979 (0.959) & 0.977 (0.969) \\ 
			Proposed (HC4m)  & 0.146 (0.219) & 0.085 (0.085) & 0.062 (0.064) & 0.953 (0.948) & 0.969 (0.947) & 0.967 (0.956) \\ 
			Proposed (HC5)   & 0.127 (9.598) & 0.082 (0.084) & 0.061 (0.065) & 0.938 (0.928) & 0.962 (0.941) & 0.962 (0.956) \\ \hline  \end{tabular}}
	\caption{Standard error and coverage probability of the 95\% confidence interval under 1:1 (2:1) stratified randomization and correct model specification for the total sample size of 30, 60, and 90.}
	\label{tab:secoverallL}
\end{table}

\begin{table}[!ht]
	\centering
	\noindent\adjustbox{max width=\textwidth}{%
		\begin{tabular}{lcccccc}
			\hline
			& \multicolumn{6}{c}{Type I error rate} \\\hline
			\multicolumn{1}{r}{Total sample size:} & 30 & 60 & 90 & 150 & 360 & 900 \\
			\multicolumn{1}{r}{Randomization ratio:} & 1:1 (2:1) & 1:1 (2:1) & 1:1 (2:1) & 1:1 (2:1) & 1:1 (2:1) & 1:1 (2:1)  \\\hline
			\multicolumn{1}{l|}{Method}& \multicolumn{6}{c}{Scenario 3: No treatment effect} \\\hline
			Proposed (model) & 0.093 (0.135) & 0.071 (0.082) & 0.065 (0.069) & 0.059 (0.060) & 0.052 (0.053) & 0.051 (0.052) \\ 
			Proposed (const) & 0.057 (0.064) & 0.039 (0.042) & 0.041 (0.042) & 0.045 (0.045) & 0.047 (0.048) & 0.049 (0.050) \\ 
			Proposed (HC0)   & 0.110 (0.168) & 0.080 (0.105) & 0.068 (0.081) & 0.059 (0.064) & 0.052 (0.055) & 0.051 (0.052) \\ 
			Proposed (HC1)   & 0.094 (0.152) & 0.071 (0.097) & 0.063 (0.075) & 0.056 (0.061) & 0.051 (0.054) & 0.050 (0.051) \\ 
			Proposed (HC2)   & 0.072 (0.124) & 0.056 (0.077) & 0.054 (0.063) & 0.052 (0.056) & 0.049 (0.052) & 0.050 (0.051) \\ 
			Proposed (HC3)   & 0.046 (0.088) & 0.038 (0.053) & 0.041 (0.047) & 0.045 (0.047) & 0.047 (0.048) & 0.049 (0.050) \\ 
			Proposed (HC4)   & 0.036 (0.073) & 0.022 (0.032) & 0.029 (0.029) & 0.039 (0.038) & 0.045 (0.046) & 0.048 (0.049) \\ 
			Proposed (HC4m)  & 0.039 (0.077) & 0.031 (0.044) & 0.035 (0.040) & 0.041 (0.044) & 0.045 (0.047) & 0.048 (0.049) \\ 
			Proposed (HC5)   & 0.060 (0.108) & 0.042 (0.054) & 0.045 (0.047) & 0.048 (0.049) & 0.048 (0.050) & 0.050 (0.050) \\  
			\hline
	\end{tabular}}
	\caption{Type I error rate at the two-sided significance level of 0.05 under 1:1 (2:1) stratified randomization.}
	\label{tab:alphaall}
\end{table}

\begin{table}[!ht]
	\centering
	\noindent\adjustbox{max width=\textwidth}{%
		\begin{tabular}{lccc|ccc}
			\hline
			& \multicolumn{3}{c}{Standard error} & \multicolumn{3}{c}{Coverage probability}\\
			\cline{2-4} \cline{5-7}
			\multicolumn{1}{r}{Total sample size:} & 150 & 360 & 900 & 150 & 360 & 900 \\
			\multicolumn{1}{r}{Randomization ratio:} & 1:1 (2:1) & 1:1 (2:1) & 1:1 (2:1) & 1:1 (2:1) & 1:1 (2:1) & 1:1 (2:1) \\\hline
			\multicolumn{1}{l|}{Method}& \multicolumn{6}{c}{Scenario 4: Model misspecification with missing covariates} \\\hline
			Proposed (model) & 0.040 (0.041) & 0.025 (0.026) & 0.016 (0.017) & 0.944 (0.949) & 0.950 (0.954) & 0.952 (0.955) \\ 
			Proposed (const) & 0.043 (0.047) & 0.026 (0.028) & 0.016 (0.017) & 0.960 (0.969) & 0.955 (0.966) & 0.952 (0.964) \\ 
			Proposed (HC0)   & 0.039 (0.040) & 0.025 (0.026) & 0.016 (0.016) & 0.941 (0.936) & 0.948 (0.946) & 0.950 (0.949) \\ 
			Proposed (HC1)   & 0.040 (0.040) & 0.025 (0.026) & 0.016 (0.016) & 0.944 (0.939) & 0.949 (0.947) & 0.950 (0.949) \\ 
			Proposed (HC2)   & 0.041 (0.042) & 0.025 (0.026) & 0.016 (0.016) & 0.950 (0.947) & 0.951 (0.950) & 0.951 (0.950) \\ 
			Proposed (HC3)   & 0.042 (0.043) & 0.026 (0.026) & 0.016 (0.016) & 0.958 (0.957) & 0.954 (0.954) & 0.953 (0.952) \\ 
			Proposed (HC4)   & 0.045 (0.047) & 0.026 (0.027) & 0.016 (0.017) & 0.968 (0.970) & 0.958 (0.959) & 0.954 (0.954) \\ 
			Proposed (HC4m)  & 0.043 (0.045) & 0.026 (0.027) & 0.016 (0.016) & 0.961 (0.961) & 0.956 (0.956) & 0.953 (0.952) \\ 
			Proposed (HC5)   & 0.042 (0.044) & 0.026 (0.026) & 0.016 (0.016) & 0.957 (0.960) & 0.954 (0.954) & 0.952 (0.951) \\ 
			\hline
			\multicolumn{1}{l|}{Method} & \multicolumn{6}{c}{Scenario 5: Model misspecification with additional unnecessary covariates} \\\hline
			Proposed (model) & 0.056 (0.057) & 0.036 (0.037) & 0.023 (0.023) & 0.946 (0.943) & 0.948 (0.948) & 0.948 (0.948) \\ 
			Proposed (const) & 0.057 (0.063) & 0.037 (0.040) & 0.023 (0.025) & 0.949 (0.964) & 0.950 (0.967) & 0.949 (0.967) \\ 
			Proposed (HC0)   & 0.056 (0.057) & 0.036 (0.037) & 0.023 (0.023) & 0.945 (0.943) & 0.948 (0.948) & 0.948 (0.948) \\ 
			Proposed (HC1)   & 0.057 (0.058) & 0.037 (0.037) & 0.023 (0.023) & 0.949 (0.946) & 0.949 (0.949) & 0.949 (0.949) \\ 
			Proposed (HC2)   & 0.057 (0.058) & 0.037 (0.037) & 0.023 (0.023) & 0.949 (0.946) & 0.949 (0.949) & 0.949 (0.949) \\ 
			Proposed (HC3)   & 0.058 (0.059) & 0.037 (0.037) & 0.023 (0.023) & 0.952 (0.949) & 0.951 (0.950) & 0.949 (0.950) \\ 
			Proposed (HC4)   & 0.057 (0.058) & 0.037 (0.037) & 0.023 (0.023) & 0.951 (0.947) & 0.950 (0.950) & 0.949 (0.949) \\ 
			Proposed (HC4m)  & 0.058 (0.059) & 0.037 (0.037) & 0.023 (0.023) & 0.953 (0.950) & 0.951 (0.950) & 0.949 (0.950) \\ 
			Proposed (HC5)   & 0.057 (0.057) & 0.036 (0.037) & 0.023 (0.023) & 0.948 (0.945) & 0.949 (0.948) & 0.949 (0.949) \\ 
			\hline
	\end{tabular}}
	\caption{Standard error and coverage probability of the 95\% confidence interval under 1:1 (2:1) stratified randomization and model misspecification for the total sample size of 150, 360 and 900.}
	\label{tab:secoverallMH}
\end{table}

\begin{table}[!ht]
	\centering
	\noindent\adjustbox{max width=\textwidth}{%
		\begin{tabular}{lccc|ccc}
			\hline
			& \multicolumn{3}{c}{Standard error} & \multicolumn{3}{c}{Coverage probability}\\
			\cline{2-4} \cline{5-7}
			\multicolumn{1}{r}{Total sample size:} & 30 & 60 & 90 & 30 & 60 & 90 \\
			\multicolumn{1}{r}{Randomization ratio:} & 1:1 (2:1) & 1:1 (2:1) & 1:1 (2:1) & 1:1 (2:1) & 1:1 (2:1) & 1:1 (2:1) \\\hline
			\multicolumn{1}{l|}{Method} & \multicolumn{6}{c}{Scenario 4: Model misspecification with missing covariates} \\\hline
			Proposed (model) & 0.123 (0.191) & 0.069 (0.072) & 0.053 (0.055) & 0.929 (0.912) & 0.939 (0.929) & 0.939 (0.941) \\ 
			Proposed (const) & 37.770 (239.092) & 0.102 (1.522) & 0.061 (0.091) & 0.951 (0.948) & 0.968 (0.967) & 0.965 (0.971) \\ 
			Proposed (HC0)   & 0.108 (0.102) & 0.067 (0.065) & 0.053 (0.052) & 0.896 (0.861) & 0.925 (0.895) & 0.933 (0.919) \\ 
			Proposed (HC1)   & 0.113 (0.107) & 0.069 (0.067) & 0.053 (0.053) & 0.909 (0.877) & 0.931 (0.902) & 0.937 (0.924) \\ 
			Proposed (HC2)   & 0.121 (0.115) & 0.073 (0.072) & 0.056 (0.056) & 0.928 (0.899) & 0.945 (0.921) & 0.947 (0.938) \\ 
			Proposed (HC3)   & 0.140 (0.139) & 0.081 (0.081) & 0.060 (0.061) & 0.952 (0.931) & 0.963 (0.943) & 0.961 (0.955) \\ 
			Proposed (HC4)   & 0.190 (134.936) & 0.097 (0.104) & 0.066 (0.071) & 0.961 (0.941) & 0.979 (0.964) & 0.976 (0.974) \\ 
			Proposed (HC4m)  & 0.151 (0.209) & 0.085 (0.086) & 0.062 (0.063) & 0.959 (0.937) & 0.970 (0.952) & 0.967 (0.962) \\ 
			Proposed (HC5)   & 0.132 (0.249) & 0.082 (0.084) & 0.060 (0.066) & 0.943 (0.919) & 0.963 (0.946) & 0.962 (0.960) \\ 
			\hline 
			\multicolumn{1}{l|}{Method} & \multicolumn{6}{c}{Scenario 5: Model misspecification with additional unnecessary covariates} \\\hline
			Proposed (model) & 0.129  (0.360) & 0.088 (0.088) & 0.072 (0.073) & 0.892 (0.908) & 0.935 (0.925) & 0.941 (0.934) \\ 
			Proposed (const) & 38.049 (1187.172) & 0.091 (0.097) & 0.074 (0.080) & 0.890 (0.916) & 0.936 (0.926) & 0.944 (0.945) \\ 
			Proposed (HC0)   & 0.119  (0.119) & 0.087 (0.088) & 0.072 (0.073) & 0.885 (0.902) & 0.932 (0.923) & 0.940 (0.933) \\ 
			Proposed (HC1)   & 0.128  (0.127) & 0.090 (0.091) & 0.074 (0.074) & 0.902 (0.924) & 0.941 (0.932) & 0.945 (0.939) \\ 
			Proposed (HC2)   & 0.132  (0.132) & 0.091 (0.092) & 0.074 (0.075) & 0.912 (0.934) & 0.944 (0.935) & 0.947 (0.940) \\ 
			Proposed (HC3)   & 0.149  (0.148) & 0.095 (0.096) & 0.076 (0.077) & 0.937 (0.957) & 0.955 (0.947) & 0.953 (0.947) \\ 
			Proposed (HC4)   & 0.166  (0.162) & 0.095 (0.095) & 0.075 (0.076) & 0.942 (0.954) & 0.954 (0.945) & 0.951 (0.944) \\ 
			Proposed (HC4m)  & 0.156  (0.155) & 0.096 (0.097) & 0.076 (0.077) & 0.944 (0.961) & 0.957 (0.949) & 0.954 (0.947) \\ 
			Proposed (HC5)   & 0.134  (0.134) & 0.091 (0.091) & 0.073 (0.074) & 0.917 (0.934) & 0.943 (0.934) & 0.946 (0.939) \\   \hline
	\end{tabular}}
	\caption{Standard error and coverage probability of the 95\% confidence interval under 1:1 (2:1) stratified randomization and model misspecification for the total sample size of 30, 60, and 90.}
	\label{tab:secoverallML}
\end{table}


\end{document}